\documentclass[12pt]{iopart}

\usepackage[dvips]{graphicx}
\usepackage{epsf}
\usepackage{epsfig}
\usepackage{amssymb}

\usepackage[latin1]{inputenc}

\begin{document}

\title[Driven Lattice Gases with Disorder and Open Boundary
Conditions]{Single-Bottleneck Approximation for Driven Lattice Gases
  with Disorder and Open Boundary Conditions}

\author{Philip Greulich$\ast$\dag\  and Andreas Schadschneider\dag\ddag}
\address{$\ast$~Fachrichtung Theoretische Physik, Universit\"at des 
Saarlandes, 66041 Saarbr\"ucken, Germany}
\address{\dag~Institut  f\"ur Theoretische  Physik, Universit\"at
zu K\"oln, 50937 K\"oln, Germany}%

\address{\ddag~Interdisziplin\"ares Zentrum f\"ur komplexe Systeme,
53117 Bonn, Germany}%

\date{\today}
\begin{abstract}
  We investigate the effects of disorder on driven lattice
  gases with open boundaries using the totally asymmetric simple
  exclusion process as a paradigmatic example. Disorder is realized by
  randomly distributed defect sites with reduced hopping rate. In
  contrast to equilibrium, even macroscopic quantities in disordered
  non-equilibrium systems depend sensitively on the defect sample.  We
  study the current as function of the entry and exit rates and the
  realization of disorder and find that it is, in leading order,
  determined by the longest stretch of consecutive defect sites
  (single-bottleneck approximation, SBA).  Using results from extreme
  value statistics the SBA allows to study ensembles with fixed defect
  density which gives accurate results, e.g.\ for the expectation
  value of the current.  Corrections to SBA come from
  effective interactions of bottlenecks close to the longest one.
  Defects close to the boundaries can be described by effective
  boundary rates and lead to shifts of the phase transitions.  Finally
  it is shown that the SBA also works for more complex models.  As an
  example we discuss a model with internal states that has been
  proposed to describe transport of the kinesin KIF1A.
\end{abstract}
\maketitle


\section{Introduction}
\label{introduction}

Driven diffusive systems play an important role in non-equilibrium
statistical physics \cite{schmittZ,gs}. Due to broken detailed
balance they allow to investigate non-equilibrium effects.
In addition, they serve as models for
various transport processes ranging from vehicular traffic
\cite{css,book} to biological transport by motor proteins
\cite{frey1,lipowsky,physlife}.  The paradigmatic model is the
\emph{totally asymmetric simple exclusion process (TASEP)} which was
first introduced to describe protein polymerization in ribosomes
\cite{mcdonald}. It was solved exactly \cite{derrida1,schuetz} which
allows to study generic properties like the phase diagram and
\emph{boundary induced phase transitions} \cite{krug1} that also occur
in more complex driven systems without resorting to approximations or
simulations.

In contrast to the homogeneous case, much less is known for the TASEP
with inhomogeneous hopping rates, i.e.\ disorder. Here in principle
one has to distinguish particle-dependent and site-dependent hopping
rates. The former case is simpler since it can be mapped onto an exactly
solvable zero-range process
\cite{evans1,krugferr,evans2,evanshanney}\footnote{For investigations
  of \emph{disordered} zero-range processes, see e.g.\ \cite{lusa}.}.
Therefore, at least for periodic boundary conditions, this case is
well understood. In contrast, in the latter case exact results are
known only for a single defect in an otherwise deterministic system
with sublattice-parallel dynamics
\cite{schuetz93a,schuetz93b,hinrichsen}.  For the general case,
several numerical and (approximate) analytical investigations
\cite{barma3,lebowitz,lebowitz2,krug} have revealed interesting
behaviour already for single defects and periodic boundary conditions.
However, far less is known for finite defect densities or open
boundary, and, especially a combination of both.

Apart from the fundamental theoretical interest in 
disorder effects in nonequilibrium systems, which are far from being
well understood \cite{stinchcombe}, these are also of direct relevance
for applications. Typical examples are found in intracellular
transport processes where molecular motors often move along 
heterogeneous tracks like DNA or mRNA (see e.g.\ \cite{kafri}).

In this work we consider systems with {\em binary} disorder, i.e.\ 
with transition rates that can take two possible values that are
randomly assigned to the sites.  Sites with a lower transition rate
are called \emph{defect sites} (or \emph{slow sites} due to their
influence on the average velocity), while those with the higher
transition rate are \emph{non-defect sites}.

For periodic boundary conditions the influence of single defect sites
has been clarified by Janowsky and Lebowitz
\cite{lebowitz,lebowitz2}.  The current through the system is limited
by the transport capacity of the defect site leading to a
density-independent current at intermediate densities. The
corresponding steady state phase-separates into a high and low-density
region separated by a shock.  In contrast, at high and low densities
the system is only affected locally by the presence of the defect.

The effects of finite defect density in a periodic system have
been studied in detail in \cite{barma,barma2,krug}, mainly numerically.
For binary distributions of defects, Tripathy and Barma \cite{barma,barma2} 
have classified two different regimes: 1) a homogeneous regime with a single
macroscopic density and non-vanishing current, 2) a segregated-density
regime, with two distinct values of density and non-vanishing current. 
Considering the partially ASEP where disorder was realized by inhomogeneous 
hopping bias, they also found a vanishing-current regime which shows 
two distinct densities, but with a current that vanishes asymptotically for
($L\to\infty$). They argue that phase separation can be understood by
a maximum current principle: For a given mean density the system
settles in a state which maximizes the stationary current. Thus the
largest stretch of slow bonds acts as current limiting segment. 
For the same system, Juh\'asz et al. \cite{juhasz1} introduced an 
effective potential and determined trapping times 
in potential wells to investigate the vanishing of the current in a 
finite-size scaling.

In the case of open boundary conditions, not only detailed balance but
also translational invariance is broken. Although in principle one
expects the same phases as in the pure system, the phase boundaries
and the nature of the transitions might change. For equilibrium
systems the Harris criterion \cite{harriscrit} allows to decide whether
critical behaviour is altered by weak disorder. For nonequilibrium
systems no such general statements are currently available 
\cite{stinchcombe}.

As in the periodic case, for a single defect site generically
disorder-induced phase separation into macroscopic regions of
different densities is observed in certain parameter regimes.  The
presence of defects leads to a decrease of the transport capacity and
the maximum current phase is enlarged compared to the pure system
\cite{kolomeiski2}.  These results have been generalized to systems
with a single stretch of consecutive defects, called
\emph{bottleneck}, or two defect sites (or bottlenecks)
\cite{chou,dong,dong2,ourpaper}.  It has been shown that for such systems
reliable analytical approximations exist which go beyond a simple
mean-field approach to take into account the relevant correlations
\cite{chou,ourpaper}.

For open systems with a single defect (or bottleneck) the current depends
on the position of the defect \cite{dong,dong2,ourpaper} if it is located
close to a boundary. This \emph{edge effect} due to the interaction between
the defect and the boundary also occurs in systems with
many defects and can be accounted for by \emph{effective boundary rates} 
\cite{ourpaper}.

The focus of most previous investigations was on individual realizations
of defect distributions while statistical properties of defect
ensembles were not considered. Numerical investigations on ensembles have
been made in \cite{derrida2}, where the influence of
defects on the phase transition between low- and high-density phase of
the randomly disordered TASEP \emph{with open boundary conditions} was
studied. It was shown that the position of this phase transition is
sensitively sample-dependent, even for large systems.
This effect is due to defects near the boundaries which
is consistent with the results in \cite{ourpaper}. Krug \cite{krug}
conjectured that also the maximum current is sample-dependent and is
mainly determined by the longest bottleneck. In \cite{dong,ourpaper} this
was shown to be correct, at least for two bottlenecks which are not
too close to each other.  These observations lead to the \emph{Single
  Bottleneck Approximation (SBA)} which is supported in this work
by numerical and analytical arguments.

For applications to real systems, macroscopic parameters and quantities
are most relevant. Since we have seen that macroscopic quantities can
depend on microscopic ones that can differ for different defect
samples, we are mainly interested in determining statistical
properties, e.g.\ probability distributions and expectation values, of
relevant quantities taking an ensemble of systems rather than looking
at single samples.  We therefor consider in this work a large ensemble
of individual finite but large systems, while the individual values of
these quantities might vary for different samples.
 
The goal of this work is to understand the phase diagrams of driven
lattice gases and give quantitative approximations for the
expectation values of critical parameter values and the maximum
current. Using Monte Carlo (MC) simulations, we therefore 
check the validity of the SBA and
the concept of effective boundary rates on individual samples not only
in the disordered TASEP, but also in a more complex model with internal
states (NOSC model without Langmuir kinetics \cite{NOSC1,NOSC2}, 
see \ref{app-NOSC}), which is a
model for intracellular transport with KIF1A motor proteins\footnote{These 
motor proteins belong to the kinesin family.}.
With the help of extreme value statistics these principles are used to 
derive approximations for expectation values (Sec.~\ref{expect val Jmax}).
After checking the accuracy of the SBA we discuss in Sec.~\ref{sec-corr}
the relevance of various possible corrections, e.g.\ through edge 
effects or effective interactions between the bottlenecks.
Finally, in Sec.~\ref{exp_PD} the influence of the disorder on the
phase diagram is investigated in more detail.


\section{Model definition}
\label{sec-def}

Although our considerations are rather general, we focus here on the
prototypical driven lattice gas, the \emph{totally asymmetric simple
  exclusion process (TASEP)}. The TASEP is defined on a lattice of $L$
sites which can either be empty or occupied by one particle. A
particle at site $j$ can move forward to its neighbour site $j+1$ if
this site is empty. The corresponding hopping rate is denoted by $p_j$. 
At the boundary sites $j=1$ and $j=L$ particles can be inserted and
removed, respectively. If site $1$ is empty a particle will be
inserted there with rate $\alpha$. On the other hand, if site $L$ is
occupied this particle will be removed with rate $\beta$.  Here we
will use a random-sequential update corresponding to continuous-time
dynamics.

In a schematic form the transition rules read
\begin{equation}
\label{TASEP rules}
\begin{array}{lll}
\mbox{Forward hopping in the bulk:} & \quad 10\to 01 & 
\mbox{with probability } p_i\Delta t \\
\mbox{Entry at the left boundary:} & \quad  0\to 1 & \mbox{with probability } 
\alpha \Delta t \\
\mbox{Exit at the right boundary}: & \quad  1\to 0 & \mbox{with probability } 
\beta \Delta t\,.
\end{array}
\end{equation}
As the indices indicate, the hopping rates $p_i$ are site dependent in
the disordered TASEP.  Other transitions are prohibited.

We will also investigate a generalization of the TASEP where each 
particle can be in two different states $1,2$. 
This model, which is defined in \ref{app-NOSC},
has been proposed to describe the dynamics of KIF1A motor
proteins on microtubules \cite{NOSC1,NOSC2}. In the NOSC model 
the forward-rebinding rate $\omega_f$ is a parameter controlling the 
average velocity of single particles for which we allow disorder.

In this paper we focus on \emph{binary disorder} for which the
hopping rates on each site can take the two possible values
$p$ and $q<p$ that are randomly distributed by the rule
\begin{equation}
p_j=
\left\lbrace
\begin{array}{l}
q \qquad\mbox{\,\, with prob. $\phi$} \\
p \qquad\mbox{\,\, with prob. $1-\phi$}
\end{array}
\right. \,.
\end{equation}
The parameter $\phi$ is the \emph{defect density}. Sites with 
reduced hopping rate $q$  will be called \emph{defect  sites}, or
\emph{slow sites}, while those 
with hopping rate $p$ are \emph{non-defect sites} or \emph{fast sites}. 
In the following we will set $p=1$ which can always be achieved by
rescaling time.

For convencience a sequence of $l$ consecutive defect sites will 
be called \emph{bottleneck of length $l$} in the following. A bottleneck
of length $l=1$ corresponds thus to an isolated slow site.

We will focus on \emph{``finite but large''} systems here, i.e.\
we neglect terms of magnitude ${\cal O}(1/L)$, but keep terms 
${\cal O}(1/\ln L)$. 
This is motivated by the facts that (a) the maximum current decreases with 
increasing bottleneck length and (b) the length of the longest bottleneck
grows logarithmically in $L$.


\section{Single Bottleneck Approximation}
\label{sec-SBA}

Besides the macroscopic structure of the stationary state in
dependence of the system parameters, i.e.\ the phase diagram,
the main focus of our investigation will be the (stationary)
maximum current $J^*$ for fixed bulk parameters $p$ and $q$:
\begin{equation}
J^* = \max_{\alpha,\beta}J(\alpha,\beta)\,.
\end{equation}
In analogy with the terminology used in traffic engineering, 
we will call $J^*$ the \emph{(transport) capacity}. 
Besides the macroscopic structure of the steady state, the
transport capacity will serve as main quantitative indicator for disorder
effects. Furthermore the critical values $\alpha_c$ and $\beta_c$ 
where the transport capacity is reached are of interest. 
In the pure TASEP one has $J^* = J(\alpha=1,\beta=1)=1/4$
and $\alpha_c = \beta_c = 1/2$. Both will change in the presence of
defects and, based on  previous results, could even be sample dependent.

Investigations in several works \cite{krug,barma,ourpaper} indicate
that in the TASEP with many defects, the longest stretch of
consecutive defects (\emph{bottleneck}) is the quantity that
contributes most to the transport capacity. This is plausible if one
assumes a local character of the bottlenecks by characterizing them
by an individual transport capacity $J^*_j(l)$ depending on the 
length $l$ and (possibly) position $j$. 
In the stationary state the total current is constant in space and 
is restricted by all bottleneck capacities, i.e.\ it can not exceed
the minimum of all $J^*_j(l)$. Since the transport capacity is decreasing 
monotonically with bottleneck size as was shown in \cite{ourpaper}, 
the minimum of $J^*_i(l)$ corresponds
to the transport capacity $J^*(l^*)$ of the longest bottleneck which
consists of $l^*$ consecutive defects.
Smaller bottlenecks do not contribute much as long as they are not too
close to the longest one \cite{ourpaper}. This motivates the
\emph{Single Bottleneck Approximation (SBA)}:
\begin{quote}
\label{SBA-quote}
The transport capacity $J^*$ of a disordered system with randomly
distributed defects is the same as the transport capacity $J^*_{\rm
  SBA}$ of a system with a single bottleneck if the length of this
bottleneck is the same as that of the longest bottleneck in the
disordered system.
\end{quote}
A similar conjecture has been made by Krug for periodic systems \cite{krug}.

The SBA reduces the problem to the much simpler one of a single
bottleneck in a system. In particular for the TASEP, efficient methods
have been developed recently, namely the {\em finite segment mean
  field theory (FSMFT)} \cite{chou} and the \emph{interaction
  subsystem approximation (ISA)} \cite{ourpaper}.  

We expect the SBA to work for generic driven lattice gases, especially
for low defect density $\phi$, where the average distance between
defects is large and their interactions can be neglected.  As an
example we have tested it not only for the TASEP, but also the
disordered NOSC model in the limit of vanishing Langmuir kinetics. In
both systems the average velocity of the particles is dependent on one
or more transition rates. In the TASEP the hopping rate $p$ is such a
parameter, while in the NOSC model the forward-rebinding rate
$\omega_f$ is a parameter controlling the average velocity.

\begin{table}
\begin{center}
\begin{tabular}{|c|c|c|c|c|c|c|c|}
\hline
$L$ & $\phi$ & $l^*$ & distance & length & $J^*_{\rm MC}$ & $J^*_{\rm SBA}$ 
    & $J^*_{\rm ISA}$ \\
\hline
1000 & 0.05 & 2 & 2  & 1 & 0.2174 & 0.2294 & 0.2229 \\
1000 & 0.1  & 3 & 12 & 1 & 0.2080 & 0.2131 & 0.2080  \\
1000 & 0.2  & 3 & 2  & 2 & 0.1963 & 0.2131 & 0.2080 \\
3000 & 0.1  & 3 & 4  & 1 & 0.2048 & 0.2131 & 0.2084  \\
3000 & 0.2  & 5 & 5  & 1 & 0.1866 & 0.1925 & 0.1901 \\
\hline
\end{tabular}
\end{center}
\caption{\label{table1}Comparison of Monte Carlo (MC) and SBA results 
  for the transport capacity $J^*$ in the disordered TASEP with different
  system sizes $L$ and defect densities $\phi$.  The transport
  capacity $J^*_{\rm MC}$ was obtained by Monte Carlo simulations for
  $\alpha=\beta=0.5$ (column 6) for fixed slow hopping rate $q=0.6$.
  This is compared with MC results ($J^*_{\rm SBA}$, column~7) and the
  results obtained by ISA \cite{ourpaper}($J^*_{\rm ISA}$, column 8) for a
  single-bottleneck system with one bottleneck in the bulk whose
  length is the same as the longest defect in the simulated disordered
  TASEP (column 3). Columns 4 and 5 give the distance and length of
  the bottleneck next to the longest one.}
\end{table}

\begin{table}
\begin{center}
\begin{tabular}{|c|c|c|c|c|c|c|c|}
\hline
$L$ & $\phi$ & $l^*$ & distance & length & $J^*_{\rm MC}$ & $J^*_{\rm SBA}$ \\
\hline
1000 & 0.05 & 2 & 4 & 1 & 0.07923 & 0.08179 \\
1000 & 0.1  & 3 & 2 & 1 & 0.07451 & 0.07643 \\
1000 & 0.2  & 6 & 3 & 1 & 0.06659 & 0.06717 \\
3000 & 0.1  & 4 & 6 & 1 & 0.07205 & 0.07213 \\
3000 & 0.2  & 6 & 3 & 1 & 0.06677 & 0.06717 \\
\hline
\end{tabular}
\end{center}
\caption{\label{table2} Same as in Table \ref{table1}, but for the 
  NOSC model without Langmuir kinetics. The forward hopping rate is
  inhomogeneous with $\omega_f^{\rm fast}\Delta t=0.58$ and $\omega_f^{\rm
    slow}\Delta t=0.32$. The other parameters are fixed: $\omega_h\Delta t=0.8, \,
  \omega_s\Delta t=0.22, \, \omega_b=0$. }
\end{table}

First we consider a fixed realization of disorder with small defect
density $\phi$. In this case we have a system with dilutely
distributed bottlenecks of different lengths. We want to test the SBA
for the disordered TASEP and the NOSC model. For this purpose we
simulated systems with different disorder samples and compared the
results for the transport capacity $J^*$ with numerical and analytical
results of systems with single bottlenecks in Table~\ref{table1}. For
each sample we identified the longest bottleneck $l^*$ and calculated
the transport capactity $J_{\rm SBA}^*(l^*)$ in a single-bottleneck
system with just one bottleneck of size $l^*$. One observes a quite
good agreement, although the SBA seems to overestimate the transport
capacity systematically. This is not surprising since effective
interactions of the bottlenecks will lead to an additional
decrease the current.  From the results in \cite{ourpaper} we expect
that the main effect comes from bottlenecks near the longest one.
There it was shown that for systems with two bottlenecks that,
although the main reduction of the transport capacity comes from the
longer one, the transport capacity will further be reduced if the
distance between the bottlenecks is small.  To illustrate this effect
we included the distance of the nearest bottleneck in
Table~\ref{table1}.  Since it is more probable to find a bottleneck
close to the longest one for larger defect density $\phi$, the results
tend to be less accurate with increasing $\phi$.

Surprisingly it seems that the values $J^*_{\rm ISA}$ obtained by the
semi-analytical ISA method \cite{ourpaper} are more accurate than the
numerical ones ($J^*_{\rm SBA}$) of the single-bottleneck system. This
is because ISA usually underestimates the value of $J^*(l)$ in the
TASEP with one bottleneck, while SBA overestimates the current. Thus
errors cancel.


\section{Probability distributions and expectation values in SBA}
\label{expect val Jmax}

As we have seen, the transport capacity depends quite strongly on the
particular sample of the defect distribution, i.e.\ the size of the
longest bottleneck. Usually in real systems the exact distribution of
defect sites is not known, particularly the size and position of the
longest defect can not be identified.
Then a statistical treatment, i.e.\ considering an ensemble of 
systems with fixed defect density, is more appropriate. It allows to
determine expectation values for quantities like 
currents and effective boundary rates (see Sec.~\ref{eff_bound_sect}).
This is especially relevant for applications e.g.\ to
intracellular transport on cell filaments. Each cell consists of a
large number of filaments that serve as tracks for motor proteins, and
often inhomogeneities play an important role \cite{kafri}.  Therefore each
filament can be modeled by a driven lattice gas on a linear
chain \cite{NOSC1,frey2} and the quantities of interest are
averages rather than the properties of individual chains.

In this section we want to approximate the expectation value of the
transport capacity $J^*(q,\phi,L)$ for fixed defect density $\phi$ and
finite but large system size $L$. In the last section we have shown that
for small $\phi$ the capacity depends approximately on the
size of the longest defect. Therefore we first determine the expectation
value for the size of the longest defect in such a system.

We now consider a given sample at defect density $\phi$ and system
size $L$. 
The $k$-th bottleneck has length $l_k$
and in the following two consecutive fast sites $j,j+1$ will be
interpreted as a bottleneck of length $l=0$ located at site $j$.
This implies that the number $N_b$ of bottlenecks is equal to
the number $N_f$ of fast sites, since each bottleneck
is followed by a fast site \footnote{We neglect the possible exception
at the right boundary.}. The bottleneck length $l$ is a random
variable with distribution
\begin{equation}
P_\phi(l)=\phi^l(1-\phi) \,.
\end{equation}
Since on average the fraction of fast sites is $(1-\phi)$, the mean
number of fast sites is $\langle N_f\rangle=(1-\phi)L$.  The length of
the longest bottleneck is $l^*=\max\lbrace l_k| k=1,...,N_f \rbrace$.
The statistics of the maximum of independently distributed random
values is governed by extreme value statistics \cite{sornette}. It says
that for a continuous probability distribution $P(l)$ that decays
exponentionally or faster for $l\to\infty$, the probability density of
$l^*$ being the maximum value of $N$ independently distributed random
values is for large $N$ asymptotically described by the \emph{Gumbel
  distribution} \cite{sornette}
\begin{equation}
G(u)=e^{-u}e^{-e^{-u}}
\label{ref-gumbel}
\end{equation}
where $u=u(l^*)$ is a rescaled and shifted function of $l^*$ depending 
on the details of the
probability distribution $P(l)$. 

However, since in our case the probability distribution is discrete we
need to be careful. Therefore, following the derivation used in
\cite{sornette} for continuous distributions, we derive the
probability distribution of the maximal bottleneck length explicitly
in order to control errors made by approximations. This will also
provide an explicit expression for $u(l^*)$.

The probability of a bottleneck being shorter than $l'$ is
\begin{eqnarray}
\label{P<}
P_{<}(l') &=& \sum_{l=0}^{l'-1} P_\phi(l)
=1-\phi^{l'}\,.
\end{eqnarray}
Since the $l_k$ are independently distributed, we have the probability
that all $l_k$ are smaller than $l'$:
\begin{eqnarray}
H_{<}(l') &=& P_{<}(l')^{N_f}=\exp\left(N_f\ln(1-\phi^{l'})\right) \nonumber\\ 
&=& \exp\left((1-\phi)L\ln(1-\phi^{l'}))\right)\,.
\end{eqnarray}
For large $L$ this probability is significantly larger than zero only for
$\phi \ll 1$ and we can use the approximation $\ln(1-\phi^{l'})\approx 
-\phi^{l'}$, thus
\begin{equation}
H_{<}(l')\approx \exp(-\phi^{l'}(1-\phi)L) \, .
\end{equation}
As was shown in \cite{sornette}, the error of this correction is 
${\cal O}(1/L^2)$ for exponential $P(l)$.
Thus we can neglect finite size corrections.

The probability that \emph{all} values are smaller than $l'$ is equal
to the probability that the maximum $l^*$ is smaller than $l'$, 
\begin{equation}
H_{<}(l')=\sum_{l^*=0}^{l'-1} {\cal P}(l^*) \,.
\end{equation}
${\cal P}(l^*)$ is the probability that the longest bottleneck
has length $l^*$ which is explicitly given by
\begin{eqnarray}
\label{P(lmax)-discrete}
{\cal P}(l^*) &=& H_{<}(l^*+1)-H_{<}(l^*) \\
&=& H_{<}'(l^*+\frac{1}{2}) + {\cal O}\left((\Delta l^*)^3\right)\nonumber\\ 
&\approx& -L (1-\phi) \phi^{l^*+\frac{1}{2}} \ln \phi \exp\left(-\phi^{l^*+
\frac{1}{2}}(1-\phi)L\right) \nonumber\\
&=& -\ln\phi\, e^{-u} e^{-e^{-u}}
=-\ln\phi\, G(u)
\label{eq-approxP}
\end{eqnarray}
where $G(u)$ is the Gumbel distribution (\ref{ref-gumbel}) and
we have introduced the function 
\begin{equation}
u(l^*)=-\left(l^*+\frac{1}{2}\right)\ln\phi-\ln(1-\phi)-\ln L\,.
\end{equation}

Now we assume this probability distribution to be continuous.
Using the Euler-Maclaurin formula the expectation value of $l^*$ becomes
\begin{eqnarray}
\langle l^* \rangle &=& \sum_{l^*}^L l^*{\cal P}(l^*)
\approx \int_{0}^{L} l^* {\cal P}(l^*) \,dl^* \nonumber\\
&=& \left. l^* H_{<}(l^*+\frac{1}{2})\right|_0^L
-\int_0^L H_{<}(l^*+\frac{1}{2}) \, dl^* 
\end{eqnarray}
since $H_{<}(l^*+1/2)$ is the cumulative distribution function of
${\cal P}(l^*)$. Therefore we have $\lim_{l^*\to\infty} H_{<}(l^*)=1$
and it is bounded. Hence
\begin{eqnarray}
\langle l^* \rangle &=& L H_{<}(L+\frac{1}{2})
-\int_{1/2}^{L+1/2} H_{<}(l^*) \, dl^* \nonumber \\
&=& L H_{<}(L+1/2) - \frac{{\rm Ei}(-L(1-\phi)\phi^{L+\frac{1}{2}})
-{\rm Ei}(-L(1-\phi)\phi^\frac{1}{2})}{\ln\phi} 
\end{eqnarray}
where ${\rm Ei}(x)$ is the exponential integral function. 
It can be expanded \cite{gradshteyn1}:
\begin{equation}
\begin{array}{ll}
{\rm Ei}(-x)=\gamma_e+\ln(x)+{\mathcal O}(x) & 
\mbox{ for small }|x|,\,x>0 \\
{\rm Ei}(-x)={\mathcal O}(e^{-x}) & \mbox{ for large }|x|,\, x>0
\end{array}
\end{equation}
where $\gamma_e\approx 0.5772$ is the \emph{Euler-Mascheroni constant}.

With these expansions we have for large $L$
\begin{eqnarray}
\langle l^* \rangle &=& L-\frac{\gamma_e+\ln(L(1-\phi)\phi^{L+\frac{1}{2}})
+{\mathcal O}(e^{-\ln\phi L})-{\mathcal O}(e^{-L(1-\phi)
\sqrt(\phi))}}{\ln \phi} \nonumber\\
&=& L-\frac{\gamma_e+\ln L+\ln(1-\phi)+\ln \phi (L+1/2)
+{\mathcal O}(\frac{1}{L})}{\ln \phi}
\end{eqnarray}
For finite but large systems, we can neglect terms of order 
${\mathcal O}(1/L)$. 
Thus we obtain for the expectation value of the longest bottleneck
\begin{equation}
\label{lmax-exp}
\langle l^* \rangle=\frac{\ln L+\ln(1-\phi)+\gamma_e}{\ln(1/\phi)}
-\frac{1}{2}\, .
\end{equation}
$\langle l^*\rangle $ diverges for infinite systems, as expected.
However, it grows only of order ${\cal O}(\ln L)$, so that we have to 
keep this term in finite but large systems.

If we approximate the transport capacity $J^*(\phi)$ for small $\phi$
by the corresponding current $J_{\rm SBA}^*$ of a system with one bottleneck,
the expectation value is given by $\langle J^*(\phi)
\rangle=\sum_{l^*=0}^\infty J_{\rm SBA}^*(l^*)\,{\cal P}(l^*)$. Due to the 
approximation by a continuous function, the norm
$\sum_{l^*=0}^\infty {\cal P}(l^*)\neq 1$ can significantly deviate
from one. In order to reduce this error we divide the result by
$\sum_{l^*=0}^\infty {\cal P}(l^*)$ 
\begin{equation}
\label{Jmax-exp}
\langle J_{\rm SBA}^* \rangle(\phi)
=\frac{\sum_{l^*=0}^\infty J_{\rm SBA}^*(l^*)\,{\cal P}(l^*)
}{\sum_{l^*=0}^\infty {\cal P}(l^*)}\,.
\end{equation}
We can now either take numerical values for $J^*(l^*)$ or (semi-)
analytical ones from \cite{ourpaper} or \cite{chou}. 
Since ${\cal P}(l^*)$ decays fast around $\langle l^* \rangle$ it 
is sufficient to take into account only few terms in (\ref{Jmax-exp}) 
in the   vicinity of $\langle l^* \rangle$.

In order to display the generic character of the SBA, we show results
for the transport capacity not only for the TASEP but also for the
disordered NOSC model without Langmuir kinetics
(Tables~\ref{table1} and \ref{table2}). We observe a good agreement in
both systems while the errors are of the same magnitude as for
individual samples.  This indicates that the probability distribution
function for the longest bottlenecks is an appropriate approximation.

\begin{table}
\begin{center}
\begin{tabular}{|c|c|c|c|c|c|c|c|}
\hline
$L$ & $\phi$ & no. of samples& $\langle J^* \rangle_{\rm MC}$ 
  & $ \langle J_{\rm SBA}^* \rangle_{\rm MC}$ \\ 
\hline
500 & 0.1 & 200 & 0.2099  & 0.2244 \\ 
1000 & 0.2 & 100 & 0.1918 & 0.2024 \\ 
3000 & 0.1 & 100 & 0.2018 & 0.2110 \\ 
3000 & 0.2 & 50 & 0.1866 & 0.1960 \\ 
\hline
\end{tabular}
\end{center}
\caption{\label{table3}Comparison of disorder averages in MC results 
and SBA results for the expectation value of the tranport capacity.
The defect hopping rate is $q=0.6$. Column~4 shows the numerical
results from MC simulations of the disordered TASEP. Column~5 displays
results by SBA using the probability distribution (\ref{eq-approxP}).
  }
\label{Jmax_av table}
\end{table}

\begin{table}
\begin{center}
\begin{tabular}{|c|c|c|c|c|c|c|c|}
\hline
$L$ & $\phi$ & no. of samples& $\langle J^* \rangle_{\rm MC}$ 
  & $ \langle J^*_{\rm SBA} \rangle_{\rm MC}$ \\ 
\hline
500 & 0.1 & 200 & 0.07495  & 0.08010 \\ 
1000 & 0.2 & 100 & 0.06852 & 0.07258 \\ 
3000 & 0.1 & 100 & 0.07438 & 0.075553 \\ 
3000 & 0.2 & 50 & 0.06852 & 0.07258 \\
\hline
\end{tabular}
\end{center}
\caption{\label{table4}Same as in Table~\ref{table3} but for the 
  NOSC model without Langmuir kinetics. The forward hopping rate is
  inhomogeneous with $\omega_f^{\rm fast}\Delta t=0.58$ and
  $\omega_f^{\rm slow}\Delta t=0.32$. The other parameters are fixed:
  $\omega_h\Delta t=0.8, \, \omega_s\Delta t=0.22, \, \omega_b=0$.  }
\end{table}


\section{Corrections to SBA}
\label{sec-corr}

In the following we consider corrections to the SBA and check the
quality of this approximation and the range of its validity by
statistical means.

In principle, corrections to the transport capacity could 
come from the following effects:
\begin{itemize}

\item The longest bottleneck (length $l^*=\max\{l_1,l_2,\cdots\}$) is
  located near the boundary, not in the bulk as assumed in SBA. Since
  the probability that a bottleneck at a given site is smaller than
  $l$ is $P_<(l)=1-\phi^l$ (see (\ref{P<})), the probability of
  finding the first longest
  bottleneck\footnote{There can be more than just one longest bottleneck.} 
of length $l$ at distance $x$ from a boundary is
  $P(x)=(1-\phi^l)^x \phi^l$. Therefore the average distance of the
  longest bottleneck can be approximated as
\begin{eqnarray}
  \langle x\rangle &\approx& \int_0^\infty x (1-\phi^{\langle
    l^*\rangle})^x\phi^{\langle l^*\rangle} \,dx=\frac{\phi^{\langle
    l^*\rangle}}{(\ln(1-\phi^{\langle l^* \rangle}))^2} \nonumber
  \\ &\approx& \phi^{-\langle l^*
    \rangle}=\phi^{1/2}L(1-\phi)e^{\gamma_e}={\mathcal O}(L)
\end{eqnarray}
where we approximated the longest bottleneck by its expectation value
(\ref{lmax-exp}).  That means for large systems the longest bottleneck
is, on average, far from the boundaries. However we see that for
\emph{finite} systems and small defect densities $\phi\ll 1$, $\langle
x \rangle$ is becoming small, so that the boundaries might affect the
transport capacity. This is due to a kind of ``degeneracy'' of the
longest bottleneck, since for small defect densities the probability
that there are \emph{many} longest bottlenecks is high (e.g.\ for
$l^*=1$ this degeneracy is ${\mathcal O}(L)$), thus contributions of
samples with a longest bottleneck near a boundary are relevant. We
therefore expect deviations from the SBA for very small defect
densities in finite systems.  The effect should vanish in the limit
$L\to\infty$ for fixed $\phi$.

\item Other smaller bottlenecks near the boundary can be treated by
introducing effective boundary rates (see next section).

\item Corrections from other bulk defects, i.e.\ ``defect-defect 
interactions''. Candidates for the leading contribution from this
type of correction would be a) other long defects, i.e.\ defects 
of length $l \leq l^*$, and b) defects (of arbitrary) 
located in the neighbourhood of the longest one.
The results of \cite{ourpaper} indicate that the second correction
is more important.

\end{itemize}

In order to estimate the corrections we consider an ensemble of
systems which all have a longest bottleneck of length $l^*$ and defect
density $\phi$. The slow hopping rate $q$ is considered to be fixed.
The longest bottleneck (or one of them in case of degeneracy) is
located at an arbitrary position and the distribution of the other
defect sites is not restricted.  For this ensemble the average
transport capacity is given by
\begin{equation}
  \label{eq1}
  \langle J^*\rangle (\phi,L,l^*) = {\sum_{\mathbf{x}}}^{\prime} 
  J_{l^*}^*(\mathbf{x})P_{\phi}(\mathbf{x})
\end{equation}
where $\mathbf{x}=(x_1,\ldots,x_N)$ denotes a defect configuration
with defects at sites $x_j$.
The sum is restricted to such configurations for which the longest
bottleneck has length $l^*$ (and therefore $N\geq l^*$). 
$P_{\phi}(\mathbf{x})$ is the probability to find the configuration
$\mathbf{x}$.

Denoting the transport capacity in SBA by $J_{SBA}$ we have
\begin{equation}
  \label{eq2}
  \langle J^*\rangle (\phi,L,l^*) = J^*_{SBA}(l^*) + {\sum_{\mathbf{x}}}^{\prime} 
\Delta J^*_{l^*}(\mathbf{x})P_{\phi}(\mathbf{x})
\end{equation}
with $\Delta J^*_{l^*}(\mathbf{x}) = J^*_{l^*}(\mathbf{x}) - J^*_{SBA}(l^*)$.
The expection value for the corrections to SBA is then
\begin{equation}
  \label{eq3}
  \langle \Delta J^*\rangle (\phi,L,l^*) = {\sum_{\mathbf{x}}}^{\prime} 
\Delta J^*_{l^*}(\mathbf{x})P_{\phi}(\mathbf{x})
 = \sum_{N^*}{\sum_{\mathbf{x}_{N^*}}}^{\prime} 
   \Delta J^*_{l^*}(\mathbf{x}_{N^*}) P_{\phi}(\mathbf{x}_{N^*}),
\end{equation}
where $N^*=N-l^*$ is the number of defects besides $l^*$ and 
$\mathbf{x}_{N^*}$ denotes the positions of these defects.
In case of a degeneracy one of the longest bottlenecks is chosen 
arbitrarily.

Since $P_{\phi}(\mathbf{x}_N)=\phi^N(1-\phi)^{L-N}=\mathcal{O}(\phi^N)$, the
leading correction in ${\mathcal O}(\phi)$ comes from
configurations with one additional defect:
\begin{equation}
  \label{eq4}
  \langle \Delta J^*\rangle (\phi,L,l^*) \approx 
  {\sum_{x_1}}^{\prime} \Delta J^*_{l^*}(x_1)P_{\phi}(x_1)
 = \left(  {\sum_{x_1}}^{\prime} \Delta J^*_{l^*}(x_1)\right)P_{\phi}(x_1)\, ,
\end{equation}
where we have used that $P_{\phi}(x_1)$ does not explicitly depend on $x_1$
(all allowed defect positions are equally probable). 

As long as the longest bottleneck is far from the boundaries, 
which we can be assumed for large systems, the transport capacity does not
depend explicitly on its position \cite{ourpaper}. Hence, instead of $x_1$
we can also use the relative position 
$d$ of the additional defect to the longest bottleneck to characterize
the configuration. If the defect is right of the longest bottleneck,
we have $d>0$, else $d<0$. Then we obtain the following necessary
condition for the SBA to work for large systems ($L\to\infty$):
\begin{equation}
  \label{eq5}
  {\sum_{d=-\infty}^\infty}^{\prime} \Delta J^*_{l^*}(d) < \infty 
\end{equation}
This condition is fullfilled if the ``bottleneck-bottleneck
interaction'' $\Delta J^*_{l^*}(d)$ decays faster than $|d|^{-1}$ for large
$|d|$, which is an restriction on the interaction strength of defects.
In \cite{ourpaper} it was shown numerically that in the TASEP this
function indeed decays faster than $|d|^{-2}$, so that (\ref{eq5}) is
fullfilled for the TASEP.

We can further quantify the contribution of the first defect near the
longest bottleneck as $P_{\phi}(x_1)=\phi(1-\phi)^{L-N}=\phi+{\mathcal
  O}(\phi^2)$. Since we have to take into account defects to the right
and the left of the longest bottleneck, we obtain in leading order
\begin{equation}
\label{firstorder}
\langle J^* \rangle(\phi,l^*)\approx\left[\,{\sum_{d}}^{\prime} 
\Delta J^*_{l^*}(d)\right]\phi \, ,
\end{equation}
where contributions with a defect on an adjacent site of the
bottleneck (i.e.\ $d=1$ and $d=-1$) do not appear in the sum, since
they belong to longer bottlenecks. Note that this approximation does
not depend on $L$.

Unfortunately currently no analytical results for $\Delta
J^*_{l^*}(d)$ are available. Therefore we have to rely on the results
of MC simulations to test the considerations made in this section. We
simulated systems with one bottleneck at a position far from the
boundaries ($>200$ sites) and one single defect for several bottleneck
lengths $l^*$ and defect position $d$ relative to the bottleneck to
obtain $J^*_{l^*}(d)$. The interaction function is then obtained as
$\Delta J^*_{l^*}(d)=J^*_{l^*}(d)-J^*_{l^*}$, where $J^*_{l^*}$ is the
transport capacity of a single bottleneck. Since $\Delta J^*_{l^*}(d)$
should decay fast with increasing $|d|$ (see also \cite{ourpaper}), it
is sufficient to take into account only defects within a finite distance
to the bottleneck\footnote{In our computations we simulated systems
  from $d=-20$ to $d=20$.}.  In order to obtain the expectation value
$\langle J^* \rangle(\phi)$ for arbitrary configurations, one has to
average over $l^*$ in the same manner as in eq.~(\ref{Jmax-exp}).

\begin{figure}[htb]
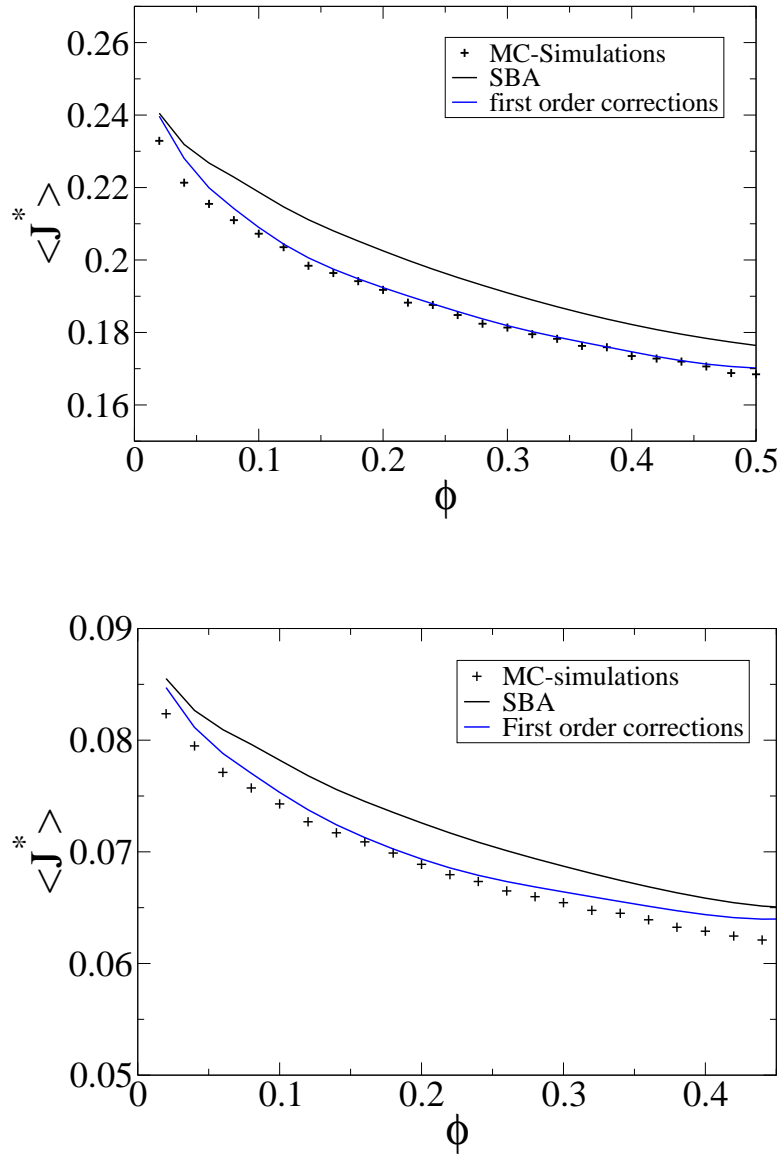

\begin{center}
  \vspace{0.8cm}
  \includegraphics[width=0.65\textwidth]{Jmax_phi_L=1000_asepdis.eps}
\end{center}
\begin{center}
  \vspace{0.9cm}
  \includegraphics[width=0.65\textwidth]{Jmax_phi_L=1000_noscdis.eps}
\end{center}
\caption{First order corrections to SBA as function of the
  defect density $\phi$ for the disordered TASEP (top) and the
  disordered NOSC model (bottom) using the probability distribution 
  (\ref{P(lmax)-discrete}) for averaging over $l^*$. The 
  slow hopping rates are $q=0.6$ in
  the TASEP and $\omega_f^{\rm slow}\Delta t=0.32$ in the NOSC model. 
  The system size is $L=1000$ in each case.}
\label{fig-SBA+corr}
\end{figure}

In Fig.~\ref{fig-SBA+corr} we have plotted average values of the
transport capacity obtained by MC simulations in dependence on the
defect density as well for the disordered TASEP and the NOSC model.
Each data point has been obtained by simulating 50 samples.  For
comparison the results in SBA and the leading order corrections
obtained by (\ref{firstorder}) are included. We see that while already
the SBA appears to be a good approximation, the accuracy of the
corrections over a wide range of defect densities is astonishing. It
comes as a surprise that in the TASEP for larger defect densities the
leading order correction, which takes into account only one additional
defect, is extremely accurate. This is not expected since for larger
$\phi$ there is a higher probability of having more than one defect in
the vicinity of the longest bottleneck.  However, these results
indicate that the position of other defects beyond the first one do
not significantly contribute to the transport capacity. Furthermore we
see that the deviation of the SBA approaches a rather constant value
for larger $\phi$, despite the factor $\phi$ in (\ref{firstorder}).
This indicates that for larger bottlenecks, the influence of single
defects on the transport capacity is weaker than for small
bottlenecks, which is consistent with results in \cite{ourpaper}. For
small defect densities configurations where the longest bottleneck is
near a boundary bottlenecks become relevant as was argued in the
beginning of the section, thus a deviation of the SBA arises in this
region, although the distance of other defects from the longest
bottleneck is large on average.

\section{The phase diagram of disordered driven lattice gases}
\label{exp_PD}

The phase diagrams of driven diffusive lattice gases that have exactly
one maximum all have the same topology. This is based on a maximum
current principle and shock dynamics \cite{gs,popkov1,diplom}.
The class of DLGs meeting this condition includes many weakly
interacting lattice gases, e.g.\ the TASEP and the NOSC model
\cite{NOSC1,NOSC2}. If disorder is included, some conceptual problems
with the expression \emph{phase diagram} arise. Usually a \emph{phase
  transition} is identified by a non-analytic behaviour of a
macroscopic quantity. In driven lattice gases these can be
discontinuities in the density (first order transitions) or kinks in
the dependence of the current on the system parameters (second order
transition). Strictly speaking these transitions only occur in
infinite systems, since non-analyticities can only in the 
\emph{thermodynamic limit}. In disordered systems, however, there is no
unique way of taking the limit $L\to\infty$ since this can not be done
with a fixed defect sample and, as we have seen, macroscopic quantities
like the transport capacity may be sample-dependent. Indeed, the
process of taking the thermodynamic limit has to be specified, since
it is ambiguous how the ``new defect sites'' by increasing $L$ are
included.  Enaud et al.~\cite{derrida2}, e.g.\ discussed two
possibilities of defining a limit $L\to\infty$ and showed that if this
limit is taken by including sites at the boundaries there actually is
no unique phase transition \emph{point} if exit rate $\beta$ is fixed
and $\alpha$ is varied.
For infinite systems, according to equation (\ref{lmax-exp}), the
length of the maximum bottleneck is infinite and thus the transport
capacity would be the same as the one of a pure system with hopping
rate $q$, $J^*=q/4$. In this work, however, we are explicitely
interested in ``finite but large systems'' and we are considering
ensembles, not individual samples. Since the longest bottleneck
increases as ${\mathcal O}(\ln L)$, the transport capacity approaches
its asymptotic value only logarithmically: $J^*(L)=q/4+{\mathcal
  O}(1/\ln L)$ (see also \cite{krug}). For finite but large systems we
have to take into account terms of the order ${\mathcal O}(1/\ln L)$.
Hence in this view, we want to consider an explicit dependence on the
system size and cannot take the thermodynamic limit to obtain phase
transitions.  In \cite{ourpaper} it is shown that if a single
bottleneck is near a boundary, phase separation cannot occur. In this
case the character of phase transitions is different, since the
current is not limited by the bottleneck anymore but by the bulk
exclusion like in the pure system. In this case the phase transition
is of second order. On the other hand, if the bottleneck is far from
the boundaries at a distance $d={\mathcal O}(L)$ there is not only a
sharp kink, but also macroscpic phase separation occurs accompanied
by a steep increase of the average density, indicating a first order
transition.

In Sec.~\ref{sec-corr} we have seen that the average distance of the
longest bottleneck from the boundaries is ${\mathcal O}(L)$. Hence on
average we have a sharp transition for large $L$. Therefore we call
this a phase transition for finite but large systems at the critical
point $\alpha_c$ where the current reaches $J^*$, although this point
depends on the system size. 


\subsection{Effective boundary rates} 
\label{eff_bound_sect}

Investigations in a TASEP with one and two bottlenecks far from the
boundaries (distance ${\mathcal O}(L)$) \cite{ourpaper,dong}
showed that the transport capacity only depends on the longer
bottleneck, while outside of the maximum current phase the current
only depends on the position of a bottleneck that is near a boundary.
This \emph{(negative) edge effect} is considerable for defects not
more than $\sim$ 20 sites away from the boundaries. The observation of
this effect motivates the concept of \emph{effective boundary rates}:
If the system is not in the maximum current phase, it can be treated
as a pure TASEP with effective boundary rates $\alpha_{\rm
  eff},\beta_{\rm eff}$ that depend on the distance and size of a
bottleneck from the boundary and differ from the real boundary rates
$\alpha$ and $\beta$.  The concept was tested in \cite{ourpaper} for
single bottlenecks and yielded good results.  The transition from low-
to high-density phase was shown to be at the line $\alpha_{\rm
  eff}=\beta_{\rm eff}$ which in general does not correspond to the
diagonal $\alpha=\beta$ in the phase diagram. The observations of
Enaud et al.~\cite{derrida2} in the disordered TASEP for different
defect samples indicate that the concept of effective boundary rates
can also be applied for the disordered TASEP.

\begin{figure}[htb]
\begin{center}
\vspace{0.75cm}
\includegraphics[width=0.65\textwidth]{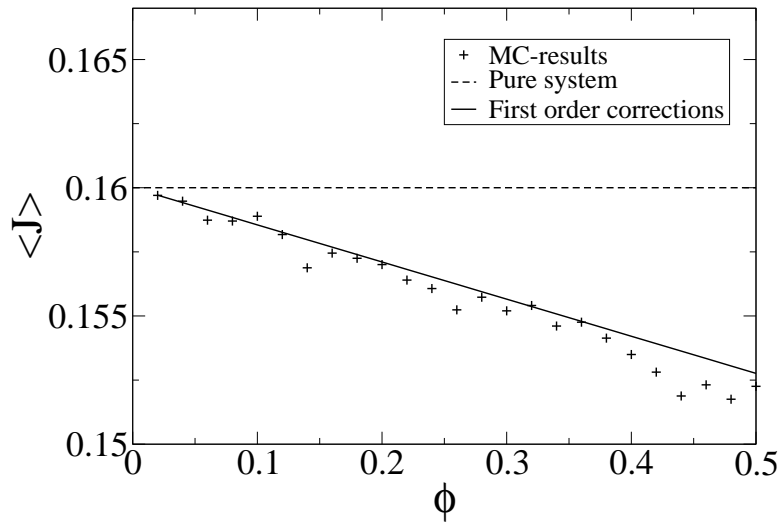}
\end{center}
\caption{Disorder average of the current in dependence of the defect 
  density in the disordered TASEP with $q=0.6$, $\alpha=0.2$ (i.e.
  $J<J^*$). MC simulations are compared with leading order corrections
  to the pure current obtained by (\ref{aeff-corr}).}
\label{J(phi)_a=0.2}
\end{figure}

Taking into account defects near the boundary, we can write the
current in the form $J(\alpha)=\alpha(1-\alpha)+\Delta J_{\alpha}({\mathbf
  x})$. Here $\alpha$ is the entry rate in the low density phase.
However due to particle-hole symmetry\footnote{Note that for
  individual defect samples, particle-hole symmetry is broken, but for
  large ensembles it is restored.} we can transfer this result to
$\beta$ and the high density phase. The defect configuration
$(x_1,x_2,...)$ is defined in the same manner as in
Sec.~\ref{sec-corr}. Indeed, taking the expectation value we can
proceed analog as in the last section to obtain the average
corrections in leading order
\begin{equation}
\label{aeff-corr}
\langle \Delta J_{\alpha} \rangle(\phi)\approx \phi \left[\sum_{d_1} 
\Delta J_{\alpha}(d_1)\right]
\end{equation}
where $d_1$ is the position of the first defect and $\Delta
J_{\alpha}(d_1)=J_{\alpha}(d_1)-\alpha(1-\alpha)$.  Thus the corrections by
defects near the boundaries are of the same magnitude as the
corrections to SBA, while the ``defect-boundary interaction'' $\Delta
J_{\alpha}(d_1)$ is in general not the same as the ``defect-defect
interaction'' $\Delta J_{l^*}(d_1)$.
Fig.~\ref{J(phi)_a=0.2} shows that results obtained from
(\ref{aeff-corr}) yield an accurate approximation for the expectation
value of the current for low entry rates.

The expectation value of the effective entry rate can then be obtained
if the current density relation of the pure system $J(\rho)$ is known.

If the relations $\alpha_{\rm eff}(\alpha),\,\, \beta_{\rm eff}(\beta)$ and
their inverses $\alpha^{-1}(\alpha_{\rm eff})$ and
$\beta^{-1}(\beta_{\rm eff})$ are known as well as the transport capacity
$J^*$, we are in principle able to map the problem of determining the
phase diagram of a disordered system on a pure system with a known
dependence of the current on the boundary rates $J(\alpha,\beta)$:

\begin{enumerate}

\item If the system current $J(\alpha_{\rm eff},\beta_{\rm eff})<J^*$ the
  system globally has the same properties as the pure one if one
  replaces the real boundary rates by the effective ones. 

\item At the points in the $\alpha-\beta$-space where
  $J(\alpha_{\rm eff},\beta_{\rm eff})=J^*$, a phase transition occurs to a
  phase separated phase occurs in which the current is independent on
  the boundary rates and maximal.
\end{enumerate}
In particular in the TASEP we can determine the expectation value of
the effective boundary rates
\begin{equation}
\label{aeff-equ}
\langle \alpha_{\rm eff} \rangle=\frac{1}{2}-\sqrt{\frac{1}{4}-
\langle J_\alpha \rangle}
\end{equation}

There is a phase transition from low density to high density phase for
$\alpha_{\rm eff}(\alpha')=\beta_{\rm eff}(\beta')\Leftrightarrow
\alpha'=\alpha_{\rm eff}^{-1}(\beta_{\rm eff}(\beta))$ which in
general is not on the diagonal $\alpha=\beta$. Nonetheless we have
\emph{on average} due to particle-hole symmetry $\langle J_\alpha
\rangle=\langle J_\beta \rangle$ that leads to $\alpha'=\beta'$ on
average. The transition to the phase separated phase is determined by
$\alpha_c(1-\alpha_c)=J^*$ or $\beta_c(1-\beta_c)=J^*$. Unfortunately,
we are not able to determine the functions $\alpha_{\rm
  eff}(\alpha),\beta_{\rm eff}$ explicitely, since for each
$\alpha,\beta$ we need to obtain a set of functions $\Delta
J_{\alpha,\beta}$ which requires much computational effort,
as long as no analytical results are available. Nonetheless, the
concept of effective boundary rates can be used to extract some
qualitative properties of the phase diagram, though obtaining
quantitative results is difficult.

\begin{figure}[htb]
\begin{center}
\vspace{0.8cm}
\includegraphics[width=0.65\textwidth]{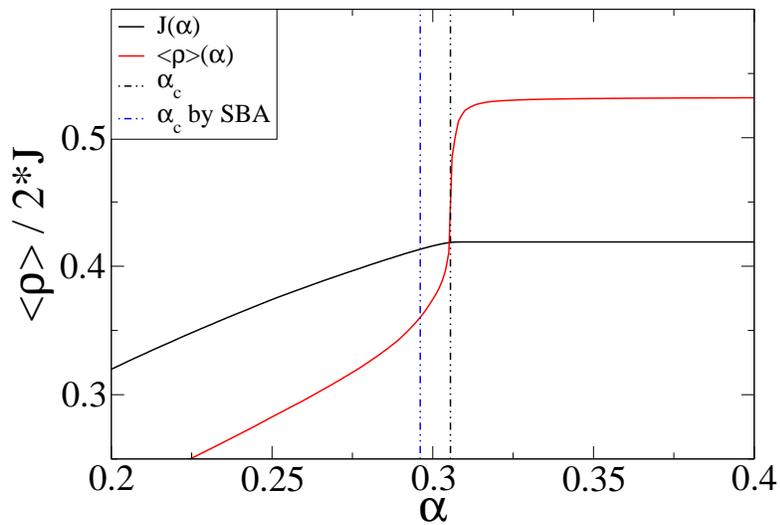}
\end{center}
\caption{Mean density $\langle \rho \rangle$ and current $J$ in dependence
  of $\alpha$ for fixed $\beta=0.9$, $q=0.6$ and $\phi=0.05$ obtained
  by MC simulation of a system with $L=2000$ and fixed defect sample.
  One observes a steep increase in at the same point where the current
  reaches the plateau.}
\label{mass+J(a)}
\end{figure}

However, since corrections of the SBA are of the same order as
corrections to the boundary rates, we can approximate $\alpha_{\rm
  eff}\approx \alpha$ and $\beta_{\rm eff}\approx \beta$ in order to
find $\alpha_c$ and $\beta_c$.  In Fig.~\ref{mass+J(a)} we plotted the
current and the average density in dependence on the entry rate
$\alpha$. One observes a steep increase in the average density at the
point where the plateau begins. Thus we can can characterize this
transition as a first order phase transition.

\begin{figure}[htb]
\begin{center}
  \vspace{0.8cm} \includegraphics[width=0.7\textwidth]{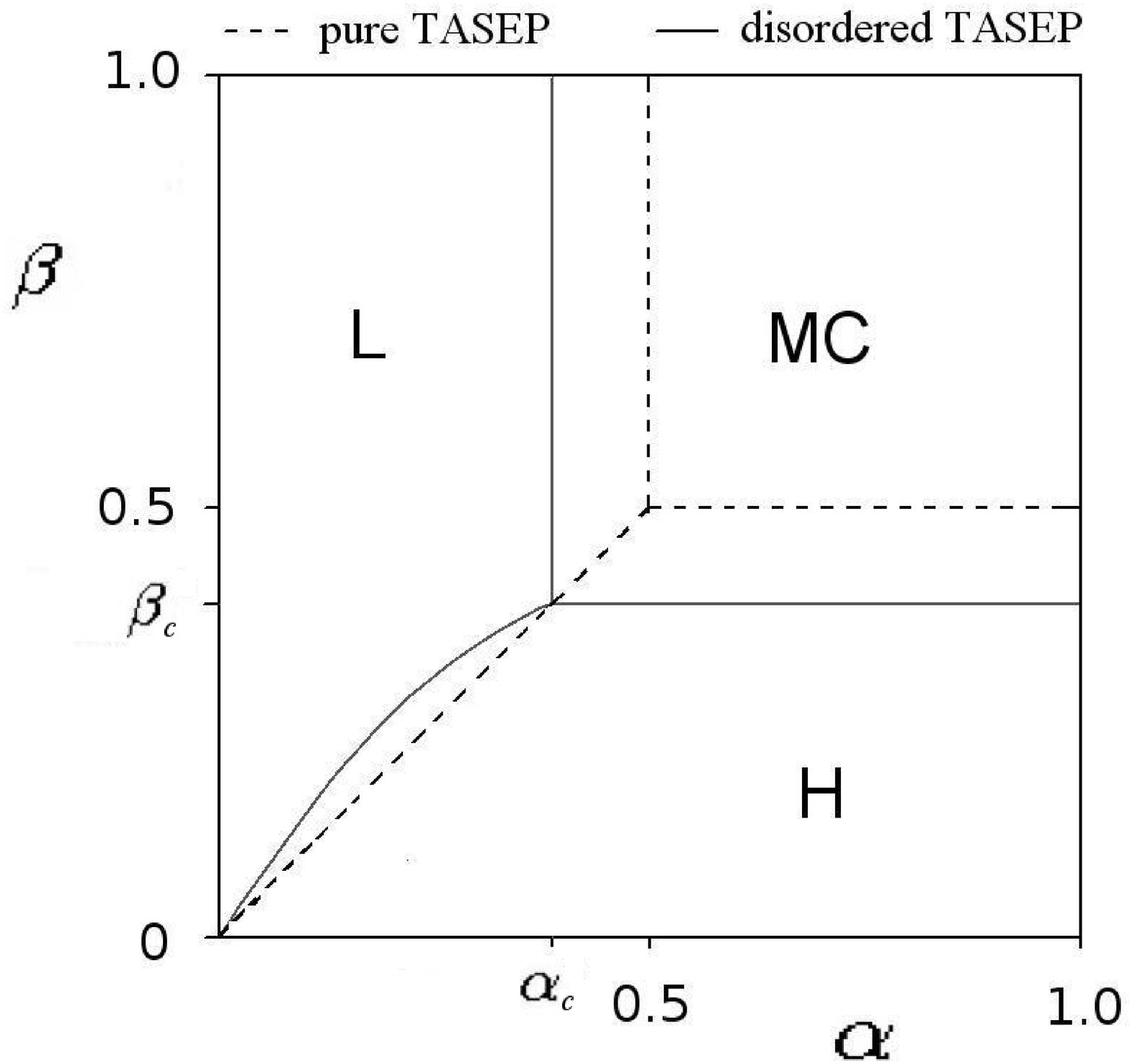}
\end{center}
\caption{Schematic phase diagram of the disordered TASEP 
  for a single defect sample. It is obtained by taking the phase
  diagram of the TASEP in dependence of $\alpha_{\rm eff}$ and
  $\beta_{\rm eff}$ with one bottleneck that corresponds to the
  longest one and rescale the axis by $\alpha_{\rm
    eff}\to\alpha,\,\beta_{\rm eff}\to\beta$. For comparison we
  included the phase transitions of the homogeneous TASEP with $p=1$
  (dashed lines).}
\label{PD D-TASEP}
\end{figure}

Fig.~\ref{PD D-TASEP} displays a sketch of the phase diagram of an
individual defect sample in the disordered TASEP. The transition line
between HD and LD is distorted compared to the homogeneous case.
Taking the disorder average, the transitions are again on the
diagonal line $\alpha=\beta$. The
maximum current phase is enlarged and can be characterized as a first
order transition compared with the pure system, since a jump of the
average density occurs at this line.


\section{Discussion and Outlook}

We have investigated disorder effects in driven lattice gases using
the TASEP with binary hopping rates as paradigmatic example.  Our
results indicate that good approximations for the expectation value of
the transport capacity (and other quantities) of an ensemble of driven
diffusive lattice gas systems can be obtained with low effort (no
averaging over disorder distribution).

The basic idea stems from the observation that the longest bottleneck
(consecutive string of slow sites) is the current limiting factor.
This suggests the possibility of calculating the transport capacity of
given defect samples by the Single Bottleneck Approximation (SBA).  It
allows to use known results for systems with only one bottleneck,
which are usually much better understood than the disordered ones, as
an efficient and accurate description.
With the help of extreme value statistics one obtains the probability
distribution for the longest bottleneck from which the expectation 
value of the transport capacity in the SBA can be determined.
Since for finite systems only a small range of bottleneck lengths
gives relevant contributions to the expectation value, it is
sufficient to have the results for a small number of single bottleneck
systems. So even for systems for which no analytical results are
available, one can get SBA results by once simulating a small number
of single bottleneck systems. Using these data approximations
of the transport capacity for arbitrary system size and defect density
(but fixed transition rates $p$, $q$) can be obtained.

We emphasize that the results obtained here are useful in two 
different situations: a) for a fixed realization of disorder, if the 
longest bottleneck can be identified, and b) for ensembles of systems 
with fixed density. In the first case, one can directly identify the
disordered system with the appropriate single-bottleneck case.
In the second case, which is also relevant for many realistic
scenarios, one can use the statistical description developed here to
obtain predictions for the ensemble.

The accuracy of the SBA can be systematically improved by
taking into account various corrections.
We found that for small defect densities the most important correction
is due to the first defect next to the longest bottleneck. It
can be expressed in terms of functions $\Delta J^*(x_1)$
that measure the contribution of a single defect at position $x_1$
relative to the bottleneck. 
A rather general argument indicates that the SBA is applicable
to a generic driven lattice gases if these functions decay faster 
than $d^{-1}$ with increasing distance. Indeed for both cases
studied here explicitly, the disordered TASEP and the NOSC model 
with vanishing Langmuir kinetics, the SBA yields good results and
thus we can expect it to work even for generic driven lattice gases.
Surprisingly, the leading order corrections appear to be
quite accurate in both systems also for larger defect densities
$\phi\approx0.5$. This indicates that other defects than the first one
only have very small influence on the transport capacity if large
bottlenecks are present\footnote{The average length of the longest
bottleneck increases with increasing defect density.}.
However for finite systems at small defect densities deviations from
the SBA occur that cannot be explained by defects near the longest
bottleneck. Here interactions of the boundaries with (one of) the longest
bottleneck can not be neglected which lead to relevant deviations
(see Sec.~\ref{sec-corr}).

We observed that deviations from the current of the pure system also
occur if the current is less than the transport capacity.  However,
these deviations are smaller in magnitude.  This effect is due to
defects near the boundaries, which was already shown before in systems
with single bottlenecks \cite{ourpaper,dong}. They can be treated in
the same manner as corrections to the SBA and we see that in this case
the first defect near the boundary is the most relevant contribution
as expected from the results before. The effect can be encompassed in
terms of effective boundary rates. In principle for known relations
between boundary rates and transport capacity, the problem of
determining the phase diagram can be mapped on a pure system using
these quantities instead of the ones of the pure system\footnote{One
  still has to be careful since the characteristics of the phases can be
  different in the disordered system, although the topology is the
  same.}. Though usually it is difficult to determine effective
boundary rates explicitely that concept is useful to obtain
qualitative properties of the phase diagram.

From a theoretical point of view the SBA and its corrections as well
as the effective boundary rates are interesting since by these
concepts disordered systems can be treated in terms of systems with
single bottlenecks and two-bottleneck systems. These are much easier
to investigate since one has to consider fixed defect configurations.
This follows the tradition of statistical physics since microscopic
properties of particles as well as particle-particle interactions
(here ``bottleneck-bottleneck interactions'' in form of the functions
$\Delta J(x)$) are used to obtain macroscopic quantities using
statistics. The concept presented in this work is rather generic
provided that microscopic properties can be obtained. This can for
example be done by numerical simulations.

Driven diffusive systems are used as models for active intracellular
transport processes. These are characterized by the directed motion of
motor proteins on microtubules. However, usually the microtubuli are
not homogeneous, but there are other microtubule associated proteins
(MAPs) that are attached to the microtubules and can form obstacles
that correspond to defects on the modelling level, impeding forward
movement of motor proteins. One example is the aggregation of tau
proteins in neurons of organisms suffering of Alzheimer's disease
\cite{mandelkov}. Furthermore there are experiments that show that
modified kinesin molecules can immobilize moving kinesins which thus
form obstacles on the microtubule track \cite{boehm}. For living
organisms the current of transported objects is a measure for the
performance of the transport system, which may not fall below a
threshold for maintaining cell metabolism and enable cell division.
Hence, the maximum current is a measure for the \emph{transport
  capacity} of a microtubule. Since binding and unbinding of molecules
to microtubules and kinesin occurs stochasticly depending on
temperature and concentration, this system meets the criterion of a
randomly disordered system. In a living organism there can be
trillions of microtubules, thus the expectation value of the maximum
current is a crucial quantity. The defect density rather than the
individual sample of defects on a microtubule is a measurable quantity
determined by the concentration of defect molecules and
temperature.

Nonetheless systems with particle conservation in the bulk are not
sufficient to serve as models for intracellular transport. One crucial
property intracellular transport exhibits is the attachment and
detachment of motor proteins. That means that one has to include these
effects in the models allowing particle creation and annihilation that
leads to a spatially varying current. The so called PFF model
\cite{frey1} includes Langmuir kinetics to the TASEP and virtually
takes into account the attachment and detachment of moter proteins. In
\cite{frey2} this model was investigated with one defect site.
The NOSC model in its original form \cite{NOSC1} also includes
creation and annihilation of particles and was used to model the
dynamics of the KIF1A motor protein using some kind of Brownian
ratchet to perform directed movement. The success of the SBA provoces
the assumption that the defects locally impose a maximum transport
capacity, so that the spatial varying current may not exceed the
transport capacity at any point. This problem is currently under
investigation and the results may help to improve our understanding of
intracellular transport with particle creation and annihilation
in the presence of defects or disorder.

\begin{appendix}

\section{The NOSC model}
\label{app-NOSC}

The NOSC model is used for modelling the dynamics of KIF1A motor
proteins on microtubules. These motor proteins can be in a strongly
bound state (1) where movement parallel to the microtubule is not
possible, and a weakly bound state (2) where it can diffuse along the
microtubule. cyclic transitions between these to states leads to a
directed net motion using a Brownian ratchet due to an asymmetric
binding potential. The transitions rules are in the bulk
\begin{equation}
\label{NOSC-bulk rules}
\fl
\begin{array}{lll}
\mbox{Transition to weakly bound state (hydrolysis):} & 1\to 2 & 
\mbox{ with probability } \omega_h \Delta t \\
\mbox{Transition to strongly bound state on site $i$:} & 2\to 1 & 
\mbox{ with probability } \omega_s \Delta t \\
\mbox{Transition to strongly bound state on site $i+1$:} & 20\to 01 & 
\mbox{ with probability } \omega_f \Delta t \\
\mbox{Diffusion to the right:} & 20\to 02 & \mbox{ with probability } 
\omega_b \Delta t \\
\mbox{Diffusion to the left:} & 02\to 20 & \mbox{ with probability } 
\omega_b \Delta t 
\end{array}
\end{equation}
at the right boundary (site 1):
\begin{equation}
\label{NOSC-left_rules}
\begin{array}{lll}
\mbox{Entry at the left boundary:} & 0\to 1 & \mbox{with probability } 
\alpha\Delta t \\ 
\mbox{Detachment:} & 1\to 0 & \mbox{never } \\ \nonumber
\mbox{Diffusion out of the system:} & 2\to 0 & \mbox{never }
\end{array}
\end{equation}
at the left boundary (site $L$):
\begin{equation}
\label{NOSC-right_rules}
\begin{array}{lll}
\mbox{Exit} & 1\to 0 & \mbox{with probability } \beta\Delta t \\ 
\mbox{Attachment} & 0\to 1 & \mbox{never} \\ \nonumber
\mbox{Diffusion out of system} & 2\to 0 & \mbox{never} 
\end{array}
\end{equation}
In the original NOSC model \cite{NOSC1} also creation and
annihilation of particles, i.e.\ Langmuir kinetics, is included,
\begin{equation}
\label{Att/Det-rules}
\begin{array}{lll}
\mbox{Creation:} & 0\to 1 & \mbox{with probability } \omega_a\Delta t \\
\mbox{Annihilation:} & 1\to 0 & \mbox{with probability } \omega_d \Delta t \\
\end{array} \,,
\end{equation}
but here we focus on the case $\omega_a=\omega_d=0$.  In our
simulations we have considered disorder in the forward-rebinding rate
$\omega_f$ which is one of the parameters that control the average
velocity of a particle. The standard parameter values are
$\omega_f^{\rm fast}=0.145$~ms$^{-1}$ for the fast rate and
$\omega_f^{\rm slow}=0.08$~ms$^{-1}$ for the slow rate. In this work we
used a timestep of $\Delta t=4$~ms, thus the transition
probabilities are obtained by four times the rates.

\end{appendix}

\section*{Acknowledgments}

This paper is dedicated to Thomas Nattermann on the occasion of his
60th birthday. We like to thank J. Krug, L. Santen and E. Frey for
helpful discussions.


\end{document}